\documentclass[11pt,moriond,epsfig]{article}
\usepackage{moriond,epsfig}

\def\Journal#1#2#3#4{{\em #1} {\bf #2}, #3 (#4).}

\def\Note#1#2{#1 (#2).}

\def\be{\begin{equation}}
\def\ee{\end{equation}}
\def\bea{\begin{eqnarray}}
\def\eea{\end{eqnarray}}

\begin{document}


\vspace*{4cm}
\boldmath
\title{CONSTRAINING $(\Omega_m,\Omega_\lambda)$ FROM STRONG LENSING}
\unboldmath
\author{ G. GOLSE, J.-P. KNEIB and G. SOUCAIL }

\address{Laboratoire d'Astrophysique, UMR 5572, Observatoire Midi-Pyr\'en\'ees\\ 
14 avenue E.-Belin, F-31400 Toulouse, France}

\maketitle\abstracts
{ The knowledge of the redshift of multiple images in cluster-lenses
allows to determine precisely the total projected mass within the
Einstein radius. The observation of various multiple images in a same
cluster is opening new possibilities to constrain the curvature of the
universe. Indeed, although the influence of $\Omega_m$ and
$\Omega_\lambda$ on the images formation is of the second order,
observations of many multiple images at different redshifts formed by
a regular cluster-lens should allow to constrain very accurately the
mass distribution of the cluster and to start to be sensitive to the
cosmological parameters entering the diameter angular distances.  We
present, analytical expressions and numerical simulations that
allow us to compute the expected error bars on the cosmological
parameters provided an HST/WFPC2 resolution image and spectroscopic
redshifts for the multiple images. Numerical tests on simulated data
confirm the rather small uncertainties we could obtain this way for the
two popular cosmological world models: $\Omega_m=0.3{\pm 0.24}$,
$\Omega_\lambda=0.7{\pm 0.5}$ or $\Omega_m=1.{\pm 0.33}$,
$\Omega_\lambda=0.{\pm 1.2}$. }

\section{Introduction}

Recent works on constraining the cosmological parameters using the CMB
and the high redshift supernovae seem to converge to a ``standard
cosmological model'' favouring a flat universe with $\Omega_m\sim 0.3$
and $\Omega_\lambda\sim 0.7$: White~\cite{White}.
 However these
results are still uncertain and depend on some physical assumptions, so 
the flat $\Omega_m=1$ model is still
possible (Le Dour {\it et al.}~\cite{LeDour}). It is therefore
important to explore other independent techniques to constrain these
cosmological parameters.

In cluster gravitational lensing, the existence of multiple images -- with known redshifts -- given by
the same source allows to calibrate in an absolute way the total
cluster mass deduced from the lens model. The great improvement in the
mass modeling of cluster-lenses that includes the cluster galaxies
halos (Kneib {\it et al.}~\cite{Kneib96}, Natarajan \& Kneib~\cite{Natarajan})
 leads to the
hope that clusters can also be used to constrain the geometry of
the Universe, through the ratio of angular size distances, which only
depends on the redshifts of the lens and the sources, and on the
cosmological parameters. The observations of cluster-lenses containing
large number of multiple images lead Link \& Pierce~\cite{Link}
(hereafter LP98) to investigate this expectation. They considered a
simple cluster potential and on-axis sources, so that images appear as
Einstein rings. The ratio of such rings is then independent of the
cluster potential and depends only on $\Omega_m$ and $\Omega_\lambda$,
assuming known redshifts for the sources. According to them, this
would allow marginal discrimination between extreme cosmological
cases. But real gravitational lens systems are more complex concerning
not only the potential but also off-axis positions of sources. They
conclude that this method is ill-suited for application to real
systems.

We have re-analyzed this problem building up on the modeling
technique developed by us. As demonstrated below, we reach a rather different
conclusion showing that it is possible to constrain $\Omega_m$ and $\Omega_\lambda$ using the
positions of multiple images at different redshifts and some physically motivated lens models.

\section{Influence of $\Omega_m$ and $\Omega_\lambda$ on the images formation}

\subsection{Angular size distances ratio term}

\begin{figure*}[h]
\psfig{figure=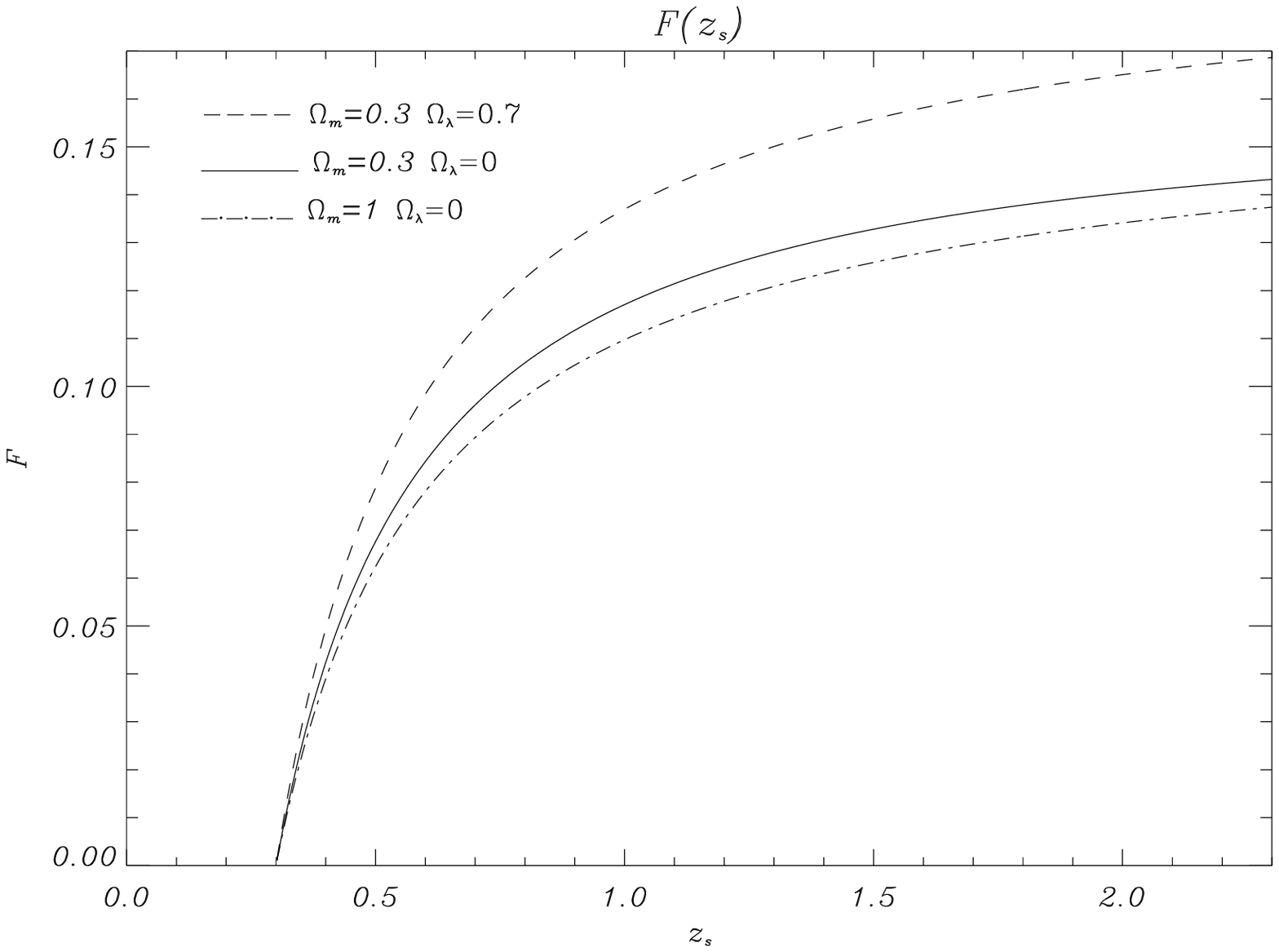, height=2.3in}
\hfill
\psfig{figure=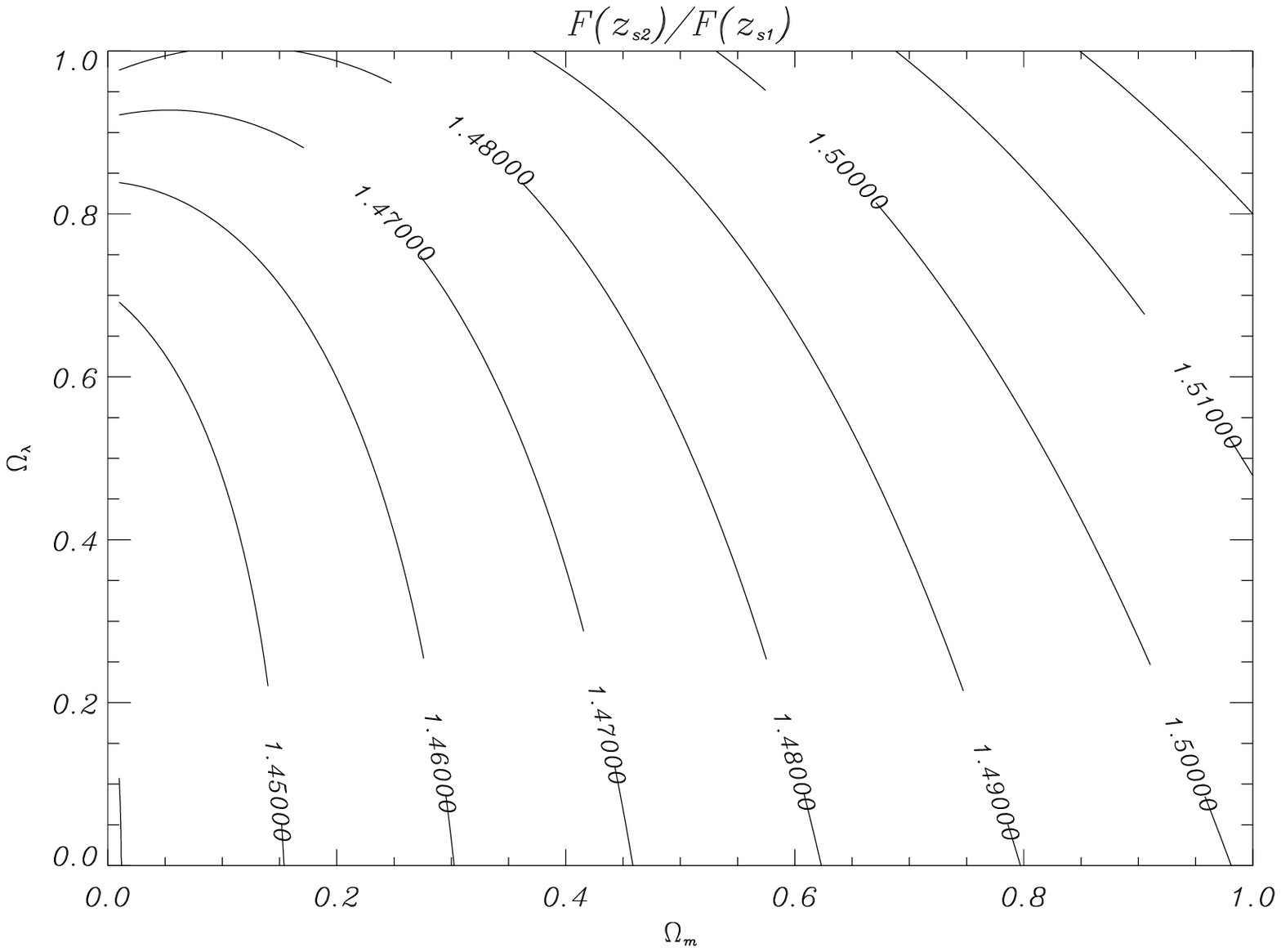, height=2.3in}
\caption{
\label{F_zs}
  Left. $F(z_s)$ for $z_l=0.3$ and various cosmological models.
  Right. $F(z_{s2})/F(z_{s1})$ as a function of $\Omega_m$ and
  $\Omega_\lambda$ for $z_l=0.3$, $z_{s1}=0.7$ and $z_{s2}=2$.  
}
\end{figure*}

In the lens equation: $\mathbf{\theta_{S}}= \mathbf{\theta_{I}} -
\displaystyle{\frac{D_{LS}}{D_{OS}}} \mathbf\nabla 
\varphi_N^{2D}(\mathbf{\theta_{I}}) $, the dependence on $\Omega_m$ and
$\Omega_\lambda$ is solely contained in the term
$F=\displaystyle{{D_{OL}}{D_{LS}}/{D_{OS}}}$.
For a given lens plane, $F(z_s)$ increases rapidly up to a certain
redshift and then stalls, with significant differences for various
values of the cosmological parameters (see Fig. \ref{F_zs}).
Thus in order to constrain the actual shape of $F(z_s)$ several
families of multiple images are needed, ideally with their redshifts
regularly distributed in $F(z_s)$ to maximize the range in the $F$
variation.

If we consider fixed redshifts for both the lens and the sources, at least 2
multiple images are needed to derive cosmological constraints. In that case
$F$ has only an influence on the modulus of
$\mathbf{\theta_{I}}-\mathbf{\theta_{S}}$. So taking the ratio of two
different $F$ terms provides the intrinsic dependence on cosmological
scenarios, independently of $H_0$.
A typical configuration leads to the Fig. \ref{F_zs} plot. The
discrepancy between the different cosmological parameters is not very
large, less than 3\% between an EdS model and a flat low matter
density one. The figure also illustrates the expected degeneracy of the method, also confirmed
by weak lensing analyzes, with a continuous distribution of
background sources ({\it e.g.} Lombardi \& Bertin~\cite{Lombardi} ).

\subsection{Relative influence of the different parameters}

We now look at the relative influence of the different
parameters, including the lens parameters, to derive expected error bars
on $\Omega_m$ and $\Omega_\lambda$. To model the potential we choose
the mass density distribution proposed by Hjorth \& Kneib~\cite{Hjorth}, 
characterized by a core radius, $a$, and a cut-off radius
$s\gg a$. We can then get the expression of the deviation angle
modulus
$D_{\theta_{I}}=\parallel\mathbf{\theta_{I}}-\mathbf{\theta_{S}}\parallel$.

For 2 families of multiple images, the relevant quantity becomes the
ratio of 2 deviation angles for 2 images $\theta_{I1}$ and
$\theta_{I2}$ belonging to 2 different families at redshifts $z_{s1}$
and $z_{s2}$. Let's define
$R_{\theta_{I1},\theta_{I2}}=\displaystyle{\frac{D_{\theta_{I1}}}{D_{\theta_{I2}}}}$. With
several families, the problem is highly constrained because a single
potential must reproduce the whole set of images. In practice we
calculate
$\displaystyle{\frac{dR_{\theta_{I1},\theta_{I2}}}{R_{\theta_{I1},\theta_{I2}}}}$
versus the different parameters it depends on. For a typical
configuration: $z_l=0.3$, $z_{s1}=0.7$,
$z_{s2}=2$, $\displaystyle{\frac{\theta_{I2}}{\theta_{I1}}}=2$,
$\displaystyle{\frac{\theta_{s}}{\theta_{a}}}=10$
($\theta_a=a/D_{OL}$,$\theta_s=s/D_{OL}$) and we assume $\Omega_m=0.3$
and $\Omega_\lambda=0.7$. We then obtain the following orders of magnitudes
for the different contributions :

\be
\displaystyle{\frac{dR_{\theta_{I1},\theta_{I2}}}{R_{\theta_{I1},\theta_{I2}}}} = 0.57\displaystyle{\frac{dz_{l}}{z_{l}}} + 0.74\displaystyle{\frac{dz_{s1}}{z_{s1}}} + 0.17\displaystyle{\frac{dz_{s2}}{z_{s2}}}
+ 0.4\left(\displaystyle{\frac{d\theta_{I1}}{\theta_{I1}}} - \displaystyle{\frac{d\theta_{I2}}{\theta_{I2}}}\right)
- 0.1\displaystyle{\frac{d\theta_{a}}{\theta_{a}}}
- 0.06\displaystyle{\frac{d\theta_{s}}{\theta_{s}}}
- 0.015\displaystyle{\frac{d\Omega_{m}}{\Omega_{m}}}
 + 0.02 \displaystyle{\frac{d\Omega_{\lambda}}{\Omega_{\lambda}}}
\ee

As expected, even with 2 families of multiple images the influence of
the cosmological parameters is of the second order. The precise value
of the redshifts is quite fundamental, therefore a spectroscopic
determination ($dz=0.001$) is essential. The position of the
(flux-weighted) centers of the images are also important. With HST
observations we assume $d\theta_I=0.1$''.

So even if the problem is less dependent on the core and cut-off
radii, they will represent the main sources of error. Taking
$d\theta_a/\theta_a= d\theta_s/\theta_s= 20$ \%, we then derive
the errors $d\Omega_m$ and $d\Omega_{\lambda}$ from the above relation
in these two cosmological scenarios :

$\Omega_m=0.3\pm0.24 $ \hspace{0.5cm} $\Omega_\lambda=0.7\pm0.5$
\hspace{1cm} or \hspace{1cm} $\Omega_m=1\pm0.33$ \hspace{0.5cm}
$\Omega_\lambda=0\pm1.2$

As confirmed by the Fig. \ref{F_zs} degeneracy plot, the method is
more sensitive to matter density than to the cosmological constant.


\section{Constraint on $(\Omega_m,\Omega_\lambda)$ from strong lensing}

\subsection{Method and algorithm for numerical simulations }

We consider basically the potential introduced in section 2.2. After considering the lens equation, fixing arbitrary
values $(\Omega_m^0$,$\Omega_\lambda^0)$ and a cluster lens redshift
$z_l$, our code can determine the images of a
source galaxy at a redshift $z_s$.
Then taking as single observables these sets of images as well as the
different redshifts, we can recover some parameters (the more
important ones being $\sigma_0$, $\theta_a$ or $\theta_s$) of the
potential we left free for each point of a grid
$(\Omega_m$,$\Omega_\lambda)$. The likelihood of the result is
obtained via a $\chi^2$-minimization, where the $\chi^2$ is computed
in the source plane. 

\subsection{Numerical simulations in a typical configuration}

To recover the parameters of the potential ( ie
$\sigma_0$, $\theta_a$, $\theta_s$ and adjusted lens parameters), we generated
 3 families of
images with regularly distributed source redshifts.

\begin{figure*}
\psfig{figure=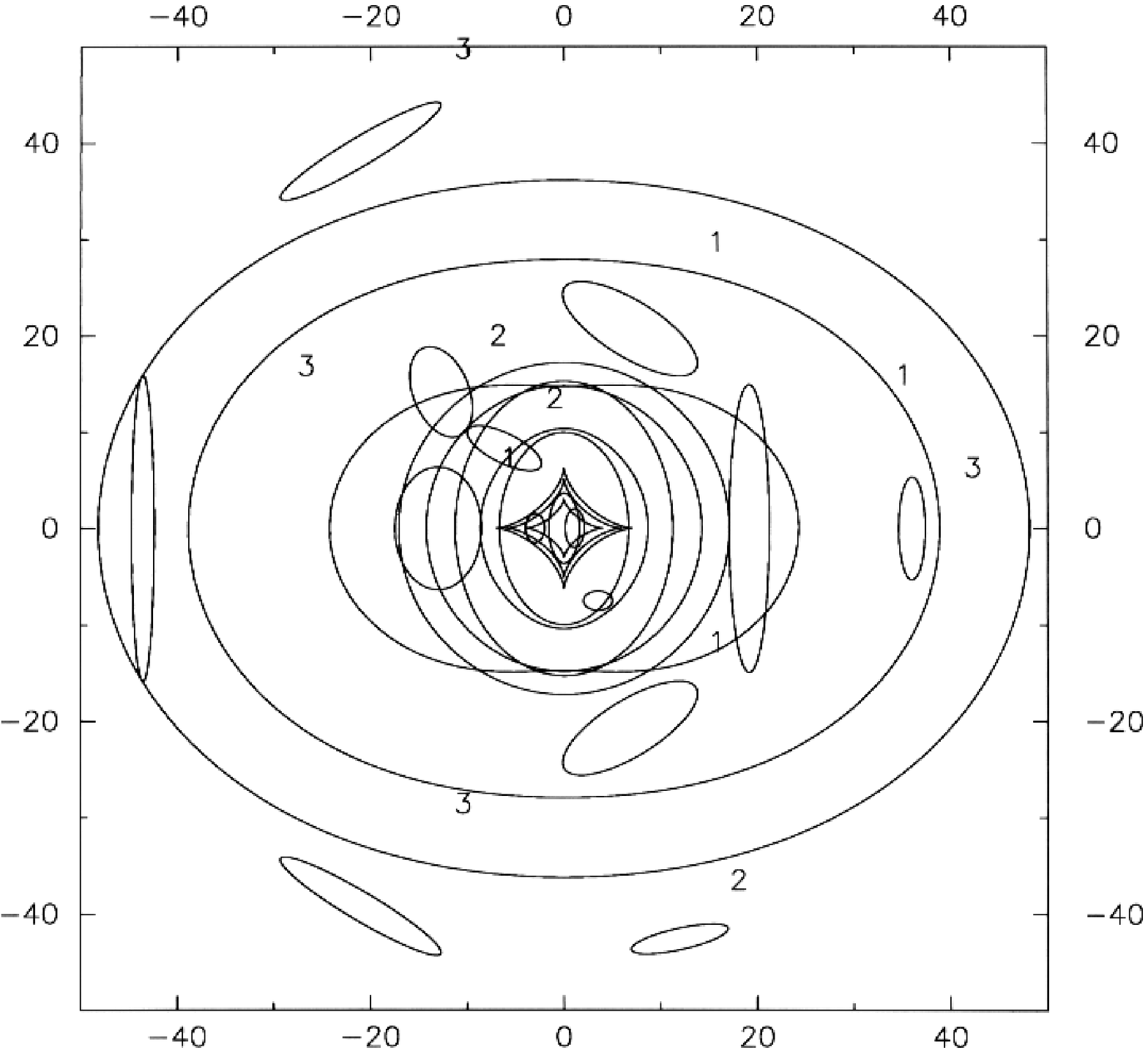, height=2.4in}
\hfill
\psfig{figure=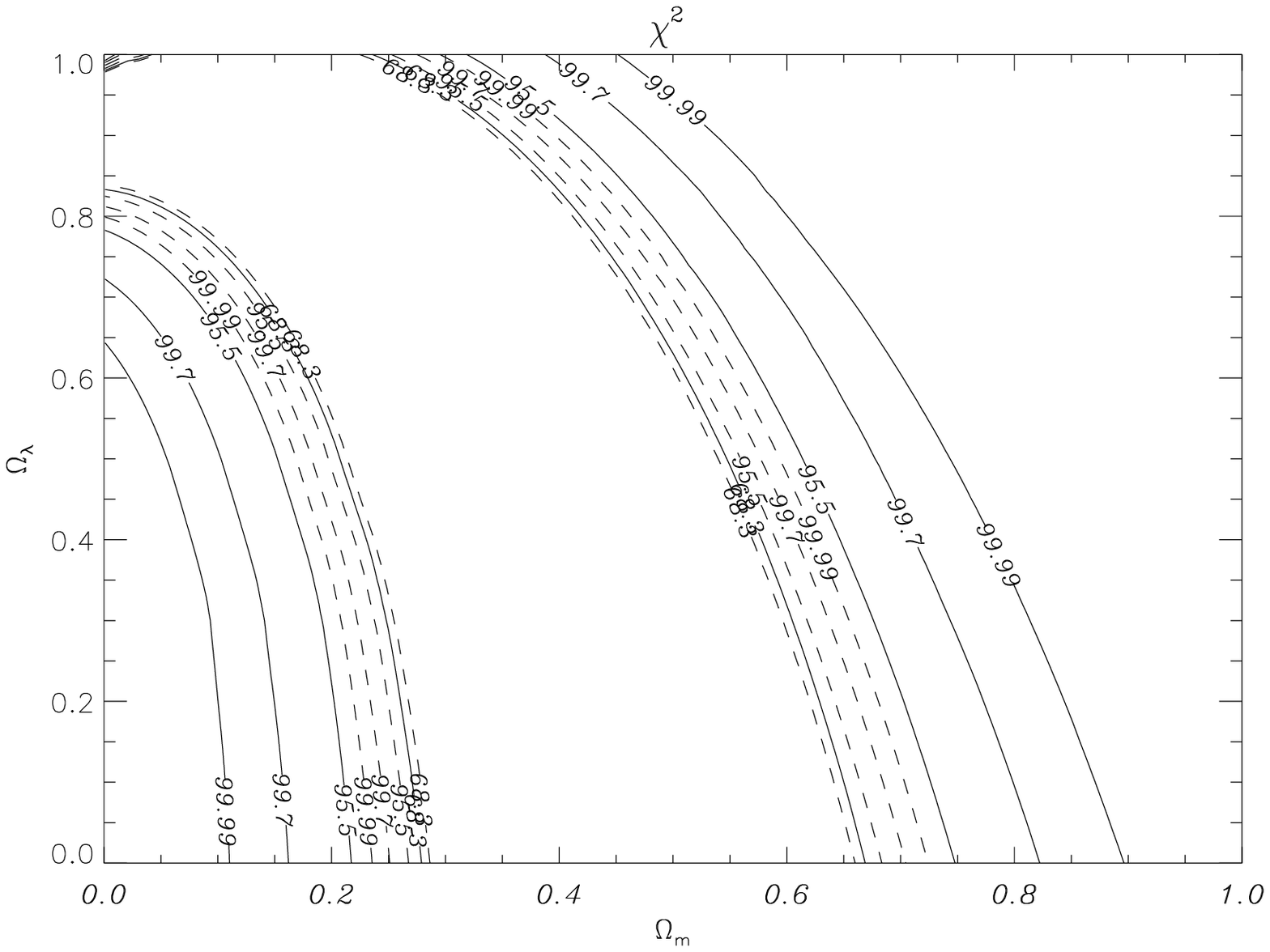, height=2.4in}
\caption{\label{3fam}
  {Left. Generation of images by a $z_l=0.3$ cluster with
  $\sigma_0=1400$ km/s, $\theta_a=13.54$'' and
  $\theta_s=145.8$''. Close to their respective critic lines, we see 3
  families of images at $z_{s1}=0.6$, $z_{s2}=1$ and $z_{s3}=2$.
  Right. Solid lines : $\chi^2(\Omega_m,\Omega_\lambda)$ confidence
  levels obtained for this configuration. Generating arbitrary values:
  $(\Omega_m^0,\Omega_\lambda^0)=(0.3,0.7)$. Dashed lines:
  $\chi^2(\Omega_m,\Omega_\lambda)$ confidence levels obtained
  considering 10 clusters in this same configuration.}}
\end{figure*}

For starting values
$(\Omega_m^0,\Omega_\lambda^0)=(0.3,0.7)$ we obtained the
Fig. \ref{3fam} confidence levels. The method puts forward a good
constraint, better on $\Omega_m$ than on $\Omega_\lambda$, and the
degeneracy is the expected one (Fig. \ref{F_zs}). Concerning the
free parameters, we also recovered in a rather good way the potential,
the variations being $\Delta\sigma_0\sim150$ km/s,
$\Delta\theta_a\sim3$''and $\Delta\theta_s\sim20$''.

This is an ``ideal'' case, of course, because we tried to recover the
same type of potential we used to generate the images, the morphology
of the cluster being quite regular and the redshift range of the
sources being wide enough to check each part of the
$F$ curve.
Such simple approach can be applied to regular clusters like
MS2137-23, which shows at least 3 families of multiple images
including a radial one. But the spectroscopic redshifts are still
missing for the moment.

\section{Conclusions \& prospects}

Following the work of LP98, we discussed a method to obtain
informations on the cosmological parameters $\Omega_m$ and
$\Omega_\lambda$ while reconstructing the lens gravitational potential
of clusters with multiple image systems at different redshifts.

This technique gives degenerate constraints, $\Omega_m$ and
$\Omega_\lambda$ being negatively correlated, with a better constraint
of the matter density. With a single cluster in a typical lensing
configuration we can expect the following error bars :
$\Omega_m=0.3{\pm 0.24}$, $\Omega_\lambda=0.7{\pm 0.5}$. To perform
that, several general conditions must be fulfilled: a cluster with a
rather regular morphology, ``numerous'' systems of multiple images, a
good spatial resolution (HST) and spectroscopic precision
for the different redshifts that should be also regularly distributed,
from $z_l$ to high values -- this requires deep spectroscopy on 8-10m class telescopes due to the
faintness of the multiple images.

Combining the study of about 10 different clusters would tighten the
error bars and lead to meaningful constraints. The dashed line
confidence levels in the Fig. \ref{3fam} are the result of a numerical
simulation made with 10 {\it identical} clusters. We are encouraged by
more and more known observations including systems with multiple
sources and we plan to apply in a first time this technique to
clusters like MS2137-23, MS0440+02, A370, AC114 and A1689.


\end{document}